\documentclass[runningheads]{llncs}
\usepackage[T1]{fontenc}
\usepackage{graphicx}
\usepackage{hyperref}
\usepackage{color}

\usepackage{easyReview}
\usepackage{rotating}
\usepackage{booktabs,threeparttable}
\begin{document}
\title{Modulation of 2-Point Discrimination Threshold Under Cheminally Induced Global Stimulation}
%
%
\author{Junjie Hua\inst{1}\orcidID{0000-0001-8918-6291} \and
Masahiro Furukawa\inst{1,2}\orcidID{0000-0003-0421-8106} \and
Taro Maeda\inst{1,2}\orcidID{0000-0002-9762-1086}}
\authorrunning{Junjie et al.}
%
\institute{Human Information Engineering Lab, Department of Bioinformatic Engineering, Graduate School of Information Science and Technology, Osaka University, Osaka 5650871, Japan \\
\email{junjie-hua@ist.osaka-u.ac.jp}
\and
Center for Information and Neural Networks, 1-4 Yamadaoka, Suita, Osaka 5650871, Japan
}
\maketitle 
\begin{abstract}
The two-point discrimination threshold (2PDT) serves as a critical indicator in the study of tactile acuity, representing the minimal distance at which an individual can differentiate two distinct points of contact on the skin. This measurement is instrumental in exploring the neural mechanisms underlying tactile perception. On the other hand, tactile acuity can be modulated by global stimulation. Prior research indicates that experimental inflammation induced by an application of capsaicin cream increases 2PDT. 
In our study, we applied chemicals (oregano, menthol, and Sichuan pepper) to selectively activate receptors that usually respond to mild physical stimuli to investigate their influence on 2PDT without inducing inflammation.
The results unveiled a pronounced augmentation of 2PDT following any form of global stimulation. Intriguingly, the cumulative effect of the chemical mix on 2PDT appeared to be additive.
These observations suggest that Wide Dynamic Range (WDR) neurons, functioning as relay nuclei with projections for touch, warmth, and cold sensations, play a pivotal role in this process. In lateral connection structures parallel to afferent nerve transmission pathways with WDR neurons as relay nuclei, global stimulation amplifies excitatory connections over inhibitory ones, thereby elevating the two-point discrimination threshold.

\end{abstract}
\section{Introduction}

Tactile acuity refers to the sharpness or sensitivity of the sense of touch. It is usually measured by the two-point discrimination threshold (2PDT), which is the smallest distance between two points that can be perceived as separate stimuli by the skin~\cite{frahm2021two}. The fingertips, palms, and forehead have very high resolution and therefore are the most sensitive areas of the body for two-point discrimination~\cite{mancini2014whole}. Other areas like the forearms, calves, and shoulder are much lesser sensitive~\cite{mancini2014whole}. This metric not only reflects the tactile acuity of the sensory system but also provides a window into the neural mechanisms underlying spatial discrimination and somatosensory processing.


On the other hand, it has been found that tactile acuity can be modulated by global stimulation. Kauppila reported that experimental inflammation induced by an application of capsaicin cream increases 2PDT~\cite{kauppila1998capsaicin}. The authors concluded that this is a result of an increase in the receptive fields of neurons in the nervous system based on the previous finding in animal experiments that capsaicin induces an increase in the size of receptive fields of the dorsal horn pain-related neurons. However, the use of just a single type of stimulation that can induce inflammation left open the possibility that combinations of stimulation that do not induce inflammation might modulate tactile acuity.

In this paper, we aim to explore the influence of global stimulation that does not induce inflammation on the two-point discrimination threshold. Evidence of nonpainful stimulation would question the involvement of pain-related neurons.
Traditional tactile devices provide temperature sensations through heat exchange with the skin~\cite{sato2016augmentation} or force feedback through pressure and vibration~\cite{chen2023sound}. These methods require contact with the skin, which means that in addition to generating the desired temperature and mechanical stimuli, there is additional pressure caused by skin contact. This additional pressure from skin contact is something we want to avoid in our experiments. Therefore, in this study, we use chemicals applied to the skin surface to activate receptors corresponding to thermal and mechanical stimuli. This method of using chemicals can avoid the additional pressure caused by skin contact.

\section{Related Work}

The process of how physical stimuli transform into perception starts with the detection of various stimuli. Thermoreceptors and mechanoreceptors are two types of sensory receptors in the body that detect changes in temperature and mechanical forces, respectively. Thermoreceptors are specialized nerve cells that are sensitive to temperature changes. The functioning of thermoreceptors is closely linked to specific ion channels, primarily transient receptor potential (TRP) channels. Different TRP channels respond to different temperature ranges.
In the range of warmth, TRPV1 is sensitive to chemical irritants (e.g., capsaicin) as well as to heat. Its heat threshold is in the vicinity of 41– 43$^{\circ}$C. Thus, activation of TRPV1 is usually accompanied by inflammation~\cite{kauppila1998capsaicin}.
Similar to TRPV1, TRPV3 is also involved in the sensation of temperature. However, it responds to lower temperatures, with a heat threshold below 39$^{\circ}$C~\cite{green2004temperature}. Thus, activation of TRPV3 hardly induces pain.
Natural monoterpenoid carvacrol from plant oregano is a known chemical irritant that specifically activates the TRPV3 channel~\cite{niu2022molecular}. 
Additionally, in the range of cold, TRPM8 was reported to have a threshold of 25 to 27$^{\circ}$C~\cite{green2004temperature}. Menthol was found to sensitize the response of TRPM8 to cold.

Mechanoreceptors are sensory receptors that respond to mechanical pressure or distortion. They are involved in sensations such as touch, pressure, and vibration. Rapidly Adapting type 1 (RA1) mechanoreceptors, also known as Meissner's corpuscles, are one of the mechanoreceptors found in the skin. They are responsible for detecting light touch and vibrations at low frequencies (about 10-50 Hz). 
It has been found that bioactive component of Sichuan pepper (also known as Szechuan pepper), hydroxy-\(\alpha\)-sanshool, activates RA1/Meissner fibers~\cite{hagura2013food}. The tingling sensation caused by Sichuan pepper on the lip does not result from pain. Instead, it results from vibration. Thus, Sichuan pepper can be used to activate the mechanoreceptor while not inducing pain.

When these receptors are activated, they send signals through peripheral nerves to the dorsal horn of the spinal cord. In the dorsal horn, primary afferent neurons synapse with second order neurons. These neurons can be unimodal, processing one type of signal, or Wide Dynamic Range (WDR) neurons, which handle multiple types of signals. The sensory signals are then relayed by these second order neurons through various ascending pathways to the thalamus and cerebral cortex. It is in the cerebral cortex, particularly the somatosensory cortex located in the parietal lobe, where this sensory information is processed and turned into conscious perception. Given the physiological findings described above, it is feasible to utilize chemicals to selectively activate receptors for perception. In fact, there has been some recent work on perception studies using chemicals. A study used capsaicin and menthol to selectively activate thermoreceptors and an illusion called Thermal Grill Illusion which is usually induced by physical warm and cold stimuli was induced~\cite{hamazaki2022chemical}. Therefore, in this study, we used chemicals applied to the skin surface to activate receptors corresponding to thermal and mechanical stimuli to investigate their influence on tactile acuity.

\section{Material and Methods}
Fifteen naive participants, aged 21 – 31 (all males), participated in all the three experiments. 

In this experiment, participants placed their right forearm on a table while receiving various global stimulations on their right volar forearm, varying in modality and intensity. The study focused on measuring the 2PDT. The sequence of modalities and intensities was randomized. For 2PDT assessment, stimuli with point-to-point distances ranging from 0 to 100 mm, increasing in 10 mm increments, were applied. A 0 mm distance indicated a single-point stimulus, applied along the forearm's proximal-distal axis. Each distance, including 0 mm, was tested twice in a randomized order. The stimulation site near the elbow was consistently positioned at the midpoint of the global stimulus's edge closest to the elbow. Participants, with eyes closed, answered the number of perceived points (1 or 2 points). Each condition involved 22 stimuli, administered at minimum 10-second intervals. To avoid influence from previous conditions, as suggested by a pilot study indicating a return to baseline after one hour, a minimum two-hour interval was maintained between different stimulus conditions.

\begin{figure}
	\centering
	\includegraphics[width=0.7\textwidth]{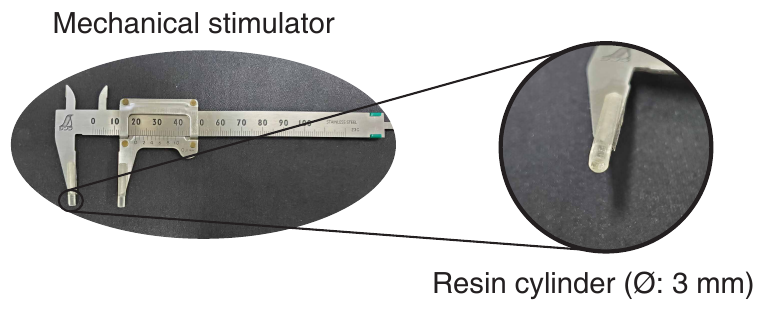}
	\caption{Mechanical stimulator used in this study. One Vernier caliber was used. The caliber had two probes attached to the jaws of the caliber. The probes were made of two resin cylinders (3 mm diameter). It was ensured that, during the stimulation, both cylinders impacted the skin simultaneously.} \label{measurement}
\end{figure}

To assess the 2PDT, a Vernier caliper was employed (refer to Fig.~\ref{measurement}). This caliper was modified to include two probes attached to its jaws, each probe consisting of a resin cylinder with a 3 mm diameter. Care was taken to ensure simultaneous contact of both cylinders with the skin during stimulation. The duration of each stimulation was set to 2 sec. Single-point stimuli were defined as the condition where the distance between the two cylinders was zero.

The 2PDT was determined for each condition individually. Prior to analysis, responses indicating one or two points were recoded from 1 or 2 to 0 or 1, respectively. The threshold was then estimated through logistic regression, fitting the data to a sigmoidal curve represented by:

\begin{equation}
	y=\frac{1}{1+e^{(a(b-x))}}
\end{equation}


In this equation, \(x\) represents the distance between points (in millimeters), \(b\) denotes the point at which \(y\) equals 0.5, and \(a\) is the slope of the curve at \(y = 0.5\). Here, \(y = 0.5\) corresponds to the separation distance at which participants perceived 50\% of the stimuli as two distinct points. During logistic regression, \(x\) and \(y\) values were derived from experimental data, while the fitting process estimated parameters \(a\) and \(b\).

\begin{sidewaystable}[ht]
	\renewcommand{\arraystretch}{1.3}
	\caption{Experimental conditions in this study}
	\label{tab:Comments}
	\centering
	\begin{threeparttable}
		
	\begin{tabular}{c|c|c|c|c|c}
		\hline
		Condition            & Exp1 & Exp2 & Exp3 & Ingredients ratio (Total volume 0.5mL) & Threshold (mm) \\ \hline
		Placebo              & $\surd$\tnote{1} & $\surd$ & $\surd$ & Salad oil  & 28.7\\ \hline
		Warm                 & $\surd$ & $\surd$ &         & Oregano oil   & 39.1\\ \hline
		Warm$\_$d            &         & $\surd$ &         & Oregano : Salad = 1 : 1  & 36.7\\ \hline
		Cold                 & $\surd$ & $\surd$ &         & Menthol oil    & 38.9\\ \hline
		Cold$\_$d            &         & $\surd$ &         & Menthol : Salad = 1 : 1   & 36.6\\ \hline
		Mechanical           & $\surd$ &         & $\surd$ & Sichuan pepper oil   & 39.7\\ \hline
		Mechanical$\_$d      &         &         & $\surd$ & Sichuan pepper : Salad = 1 : 1   & 38.5\\ \hline
		Mix1                 &         & $\surd$ & $\surd$ & Oregano : Menthol = 1 : 1  & 47.6\\ \hline
		Mix1$\_$d            &         & $\surd$ & $\surd$ & Oregano : Menthol : Salad = 1 : 1 : 2  & 39.4\\ \hline
		Mix2                 &         &         & $\surd$ & Oregano : Menthol : Sichuan pepper = 1 : 1 : 1  & 51.3\\ \hline
		Mix2$\_$d            &         &         & $\surd$ &  Oregano : Menthol : Sichuan pepper : Salad = 1 : 1 : 1 : 3   & 45.3\\ \hline
	\end{tabular}
	
	\begin{tablenotes}
		\footnotesize
		\item[1] The $\surd$ mark means the condition was used in the analysis of the corresponding experiment.
	\end{tablenotes}
	
	\end{threeparttable}
\end{sidewaystable}

\section{Experiment 1}
\subsection{Procedure}


We initially verified that the application of global chemically-induced warm, cold, and mechanical stimuli leads to an increase in the 2PDT. Salad oil was employed as a control substance. To stimulate warmth receptors, pure oregano oil (R V Essential), known for its specific activation of the TRPV3 channel~\cite{niu2022molecular}, was used. For cold receptors, menthol oil (Komamori, 30\% menthol concentration) was applied. Additionally, Sichuan pepper oil, with a hydroxy-\(\alpha\)-sanshool concentration of 4.36g/kg, was utilized to activate mechanoreceptors. Using eyedropper pipettes and cotton swabs, 0.5 mL of these oils were uniformly applied to a 40 mm x 40 mm area on the proximal forearm. The 2PDT test was performed 30 minutes following the application.


\subsection{Results of Experiment 1}

\begin{figure}
	\centering
	\includegraphics[width=0.9\textwidth]{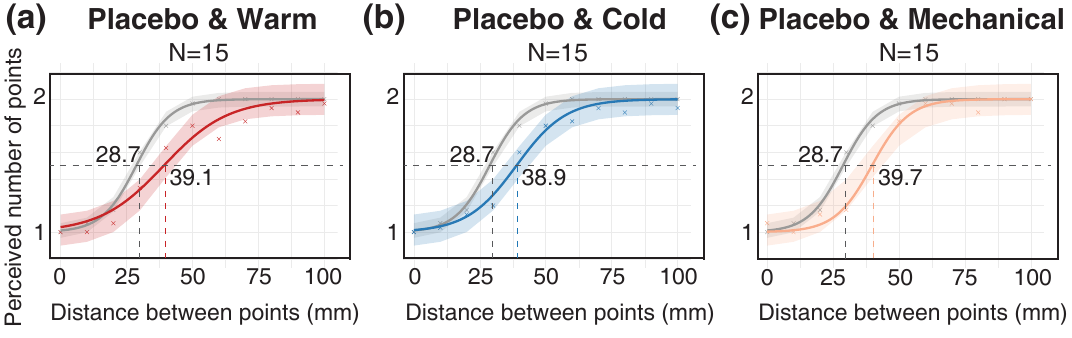}
	\caption{Results of Experiment 1. Fitted psychometric functions to the perceived number of points when globally applied with chemical-induced stimuli. The gray curve in each subplot stands for the results under the placebo condition (salad oil). The red, blue, and orange curve stands for the results under chemical-induced warm (oregano), cold (menthol), and mechanical (Sichuan pepper) stimulation.} 
	\label{result1}
\end{figure}

Each subplot in the graph represents a comparison between the control dataset and one of the other datasets (warm, cold, or mechanical) using a sigmoid function fit (see Fig.~\ref{result1}). The x-axis of each subplot stands for the distance between the two points. The y-axis of each subplot stands for the perceived number of points. The gray curve in each subplot stands for the results under the placebo condition. Compared to the placebo, chemically-induced warm, cold, and mechanical stimulation caused a rightward shift in the curve representing perceived numbers of points, indicating a reduction in sensitivity. Specifically, a greater point distance was required for participants to perceive them at an equivalent level to the placebo condition. Under warm, cold, and mechanical stimulation, the Two-Point Discrimination Threshold (2PDT) increased by 10.4 mm, 10.2 mm, and 11 mm, respectively. Additionally, this figure includes a 95$\%$ confidence interval for each condition. Throughout the experiment, no instances of pain were reported by any of the participants.

\section{Experiment 2}
\subsection{Procedure}
In experiment 1, we confirmed that the application of chemically induced stimulation (warm/cold/mechanical) enlarges 2PDT. In experiment 2, we investigated the mixture effect of warm and cold stimulation. Specifically, we tested whether the warm and cold stimulation is additive. Undiluted and 50$\%$ concentration conditions tested are listed in Tab.~\ref{tab:Comments}. Prior to the experiment, we mixed according to the proportions shown and stored the mixture in plastic bottles. The experimental methods were the same as in Experiment 1.

\begin{figure}
	\centering
	\includegraphics[width=0.8\textwidth]{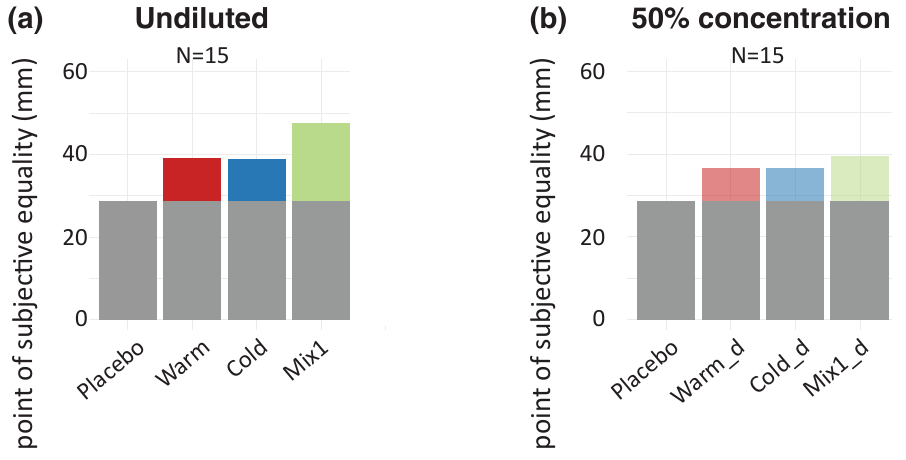}
	\caption{Results of Experiment 2. We investigated the mixture effect of chemical-induced warm and cold stimulation. The gray bar stands for the result under the placebo condition. Other bars stand for the changes due to corresponding conditions (Tab.~\ref{tab:Comments}). (a) Undiluted conditions. (b) 50$\%$ concentration conditions.} 
	\label{exp2}
\end{figure}

\subsection{Results of Experiment 2}
The point of subjective equality (PSE) of each corresponding condition in Experiment 2 is shown in Fig.~\ref{exp2} and Tab.~\ref{tab:Comments}. Results across participants were used to calculate the PSE. 
Fig.~\ref{exp2}.(a) illustrates the results under undiluted conditions.
The gray bar stands for the result under the placebo condition.
The red, blue, and green bars stand for the change due to warm, cold, and mix1 (Oregano:Menthol = 1:1) stimuli, respectively. The effect of mix1 was greater than the effect under both warm and cold conditions.
A similar relationship was observed when the stimuli were diluted (Fig.~\ref{exp2}.(b)). The effect of mix1$\_$d was greater than the effect under warm$\_$d and cold$\_$d conditions.
This result suggests that the combined effect of chemical-induced warm and cold stimuli on 2PDT appeared additive when compared to the influence of each stimulus individually.
Additionally, compared with the results under undiluted conditions, the 2PDT under the corresponding diluted conditions was smaller (Tab.~\ref{tab:Comments}), indicating that the 2PDT is influenced by concentration.
Throughout the experiment, no instances of pain were reported by any of the participants.


\section{Experiment 3}
\subsection{Procedure}

In experiment 2, we investigated the mixture effect of warm and cold stimulation. This can be seen as an interaction in the category of thermal sensation. In experiment 3, we investigated the mixture effect of thermal stimulation and mechanical stimulation. We tested whether the mixture of warm and cold and mechanical stimulation was additive. 
Undiluted and 50$\%$ concentration conditions tested are listed in Tab.~\ref{tab:Comments}. The experimental methods were the same as in Experiments 1 and 2.

\begin{figure}
	\centering
	\includegraphics[width=0.8\textwidth]{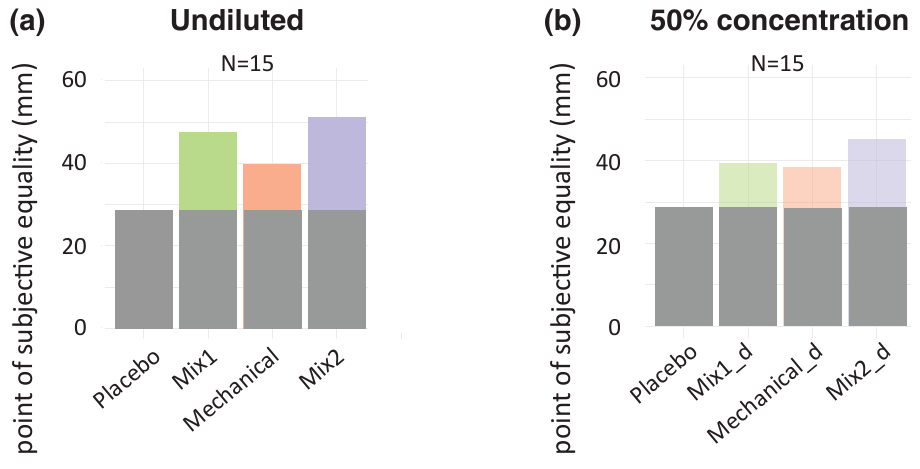}
	\caption{Results of Experiment 3. We investigated the mixture effect of chemical-induced thermal and mechanical stimulation. The gray bar stands for the result under the placebo condition. Other bars stand for the changes due to corresponding stimuli. (a) Undiluted conditions. (b) 50$\%$ concentration conditions.} 
	\label{exp3}
\end{figure}

\subsection{Results of Experiment 3}
PSE of each corresponding condition in Experiment 3 is shown in Fig.~\ref{exp3} and Tab.~\ref{tab:Comments}. Results across participants were used to calculate the PSE. 
Fig.~\ref{exp3}.(a) illustrates the results under undiluted conditions.
The purple bar was higher than the green and orange bars, indicating that the effect of mix2 (Oregano : Menthol : Sichuan pepper = 1:1:1) was greater than the effect under mix1 (Oregano:Menthol = 1:1) and mechanical conditions.
A similar relationship was observed when the stimuli were diluted (Fig.~\ref{exp3}.(b)). The effect of mix2$\_$d was greater than the effect under mix1$\_$d and mechanical$\_$d conditions.
These results suggest that the combined effect of chemical-induced thermal and mechanical stimuli on 2PDT appeared additive when compared to the influence of each stimulus individually.
Additionally, the influence of concentration was also observed.
Throughout the experiment, no instances of pain were reported by any of the participants.

\section{Discussion}

In this study, we demonstrated a modulation of the Two-Point Discrimination Threshold (2PDT) in response to global warm, cold, and mechanical stimuli, without triggering pain. To investigate the effects of these stimuli on 2PDT, while avoiding skin deformation and inadvertent mechanical stimuli that could distort the results, we applied chemicals (oregano, menthol, and Sichuan pepper) to selectively activate receptors that usually respond to mild warm, cold, and mechanical stimuli. Our results indicated a notable increase in 2PDT across all conditions—warm, cold, or mechanical stimulation—suggesting a consistent modulation of tactile acuity in the absence of inflammation. Furthermore, we observed that the combined effect of mixed chemical stimuli on 2PDT seemed to be cumulative.


It has been previously reported that capsaicin-induced inflammation globally increases the 2PDT~\cite{kauppila1998capsaicin}. In our study, we demonstrate that global application of chemically induced thermal and mechanical stimuli similarly elevates 2PDT but does so without inducing inflammation (Experiment 1). This finding indicates that tactile acuity can be modulated by multimodal stimuli independently of inflammation, suggesting that the regulatory mechanism is not mediated by nociceptive specific neurons in the nervous system. Our results imply the existence of a relay nucleus capable of processing both thermal and mechanical stimuli.

Previous research has established that Wide Dynamic Range (WDR) neurons in the spinal cord are responsive to various sensory inputs, including mechanical, warm, and cold stimuli~\cite{le2008wide,green2004temperature}. Furthermore, these neurons are known to be activated by both innocuous and noxious thermal stimuli~\cite{khasabov2001enhanced,harper2014psychophysical}. Our findings indicate that WDR neurons may serve as integral relay nuclei in the spinal cord, with projections accommodating touch, warmth, and cold stimuli. This suggests a critical role for WDR neurons in the modulation of tactile acuity.

The variation in the 2PDT was not constant but influenced by the concentration of the stimuli. Further, our data from Experiments 2 and 3 demonstrate that a combination of multiple stimuli produces a more pronounced effect than any single stimulus. Specifically, the aggregate impact of the chemical mixture on 2PDT seemed to be additive in nature compared to the individual effects of each stimulus.
These findings suggest that signals related to warm, cold, and mechanical stimuli converge on Wide Dynamic Range (WDR) neurons, and the integration of these signals might amplify the stimulation to levels typically elicited by more intense stimuli. This potent activation of WDR neurons results in a significant increase in 2PDT. Additionally, the cross-modal additivity of these signals indicates the presence of an ``intensity channel'' in the human body, based on WDR neurons~\cite{bouhassira2005investigation}. Within this channel, the specific nature of the stimulus is secondary.



\begin{figure}
	\centering
	\includegraphics[width=0.8\textwidth]{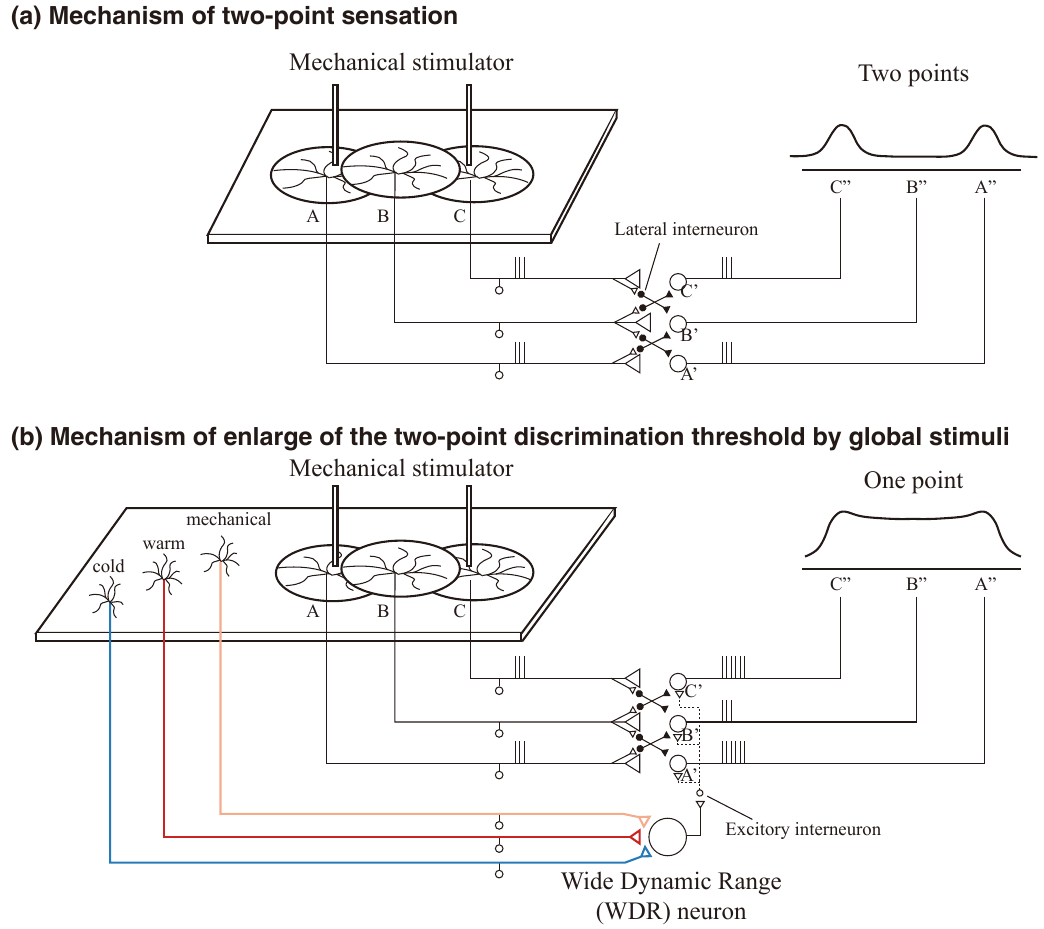}
	\caption{Mechanism of two-point discrimination sensation and mechanism of enlargement of two-point discrimination threshold by globally warm, cold, and mechanical stimulation.} 
	\label{mechanism}
\end{figure}

Here, we discuss the mechanism of two-point discrimination sensation. Mechan-oreceptors transduce mechanical deformation of the skin, induced by a mechanical stimulator, into neural signals (refer to Fig.~\ref{mechanism}.(a)). When two stimuli are applied within the same receptive field, they are perceived as a single point. However, if one stimulus falls within an adjacent receptive field, two-point discrimination becomes unfeasible, resulting in a single-point perception~\cite{maeyama1984}. Thus, for effective two-point discrimination, an unstimulated receptive field separating the two stimuli is essential. The receptor responses are relayed to the spinal cord, where lateral interneurons modulate the activity, inhibiting adjacent second order neurons (A$^{\prime}$, B$^{\prime}$, and C$^{\prime}$)~\cite{maeyama1984}. The processed signals are then conveyed to the brain, culminating in the recognition of two distinct points.

We propose a mechanism for the enlargement of the two-point discrimination threshold in response to global stimuli, encompassing warmth, cold, and mechanical factors. Upon exposure to these stimuli, the corresponding receptors are activated, transmitting signals to the spinal cord. At this juncture, Wide Dynamic Range (WDR) neurons serve as key relay nuclei~\cite{le2008wide,green2004temperature}. These neurons integrate signals of various modalities, leading to an additive effect that results in increased firing frequency of WDR neurons, compared to when a single stimulus type is received. Concurrently, excitatory interneurons amplify the activity of A$^{\prime}$, B$^{\prime}$, and C$^{\prime}$ neurons. The processed stimuli are then relayed to the brain, where they stimulate A$^{\prime \prime}$, B$^{\prime \prime}$, and C$^{\prime \prime}$ neurons. Consequently, stimuli originally perceived as two distinct points are now interpreted as a single point, thereby expanding the two-point discrimination threshold (refer to Fig.~\ref{mechanism}.(b)).

In an investigation of the two-point fused sensory area and the influence of acupuncture stimulation, previous authors observed an expansion of the fused sensory area following acupuncture~\cite{maeyama1984}. They suggested that acupuncture-induced afferent nerve impulses interfere with the lateral inhibitory mechanism through inhibitory connections. In this model, acupuncture administration leads to an expansion of the fused sensory area. Because of the inhibitory connections, second order neurons at greater distances cannot be activated. This implys that variations in acupuncture intensity (or the output of the relay nucleus) do not alter the extent of expansion. This interpretation, however, contrasts with the findings of our current study. To explain our results, we propose a model based on excitatory connections. In this framework, excitatory connections facilitate the activation of second order neurons at greater distances, resulting in a progressive enlargement of the 2PDT.

Moreover, considering that the mechanical stimulator generates mechanical stimuli, it is possible that the second order neurons (A$^{\prime}$, B$^{\prime}$, and C$^{\prime}$) in Fig.~\ref{mechanism} are WDR neurons. When these neurons receive excitatory inputs from far away due to the global application of stimulation, it may be manifested as an increase in the receptive field. This hypothesis aligns with the observed increase in the receptive fields of WDR neurons following capsaicin application, as reported in a prior study~\cite{kauppila1998capsaicin}.

\section{Conclusion}
It was shown that chemical-induced warm, cold, and mechanical stimuli lead to an increase in the two-point discrimination threshold (2PDT) without inducing inflammation. Moreover, the combined effect of chemical-induced stimuli on 2PDT appeared to be additive. 
These findings suggest that Wide Dynamic Range (WDR) neurons, which act as relay nuclei with projections for touch, warmth, and cold sensations, play a crucial role in this process. In lateral connection structures parallel to afferent nerve transmission pathways with WDR neurons as relay nuclei, global stimulation amplifies excitatory connections over inhibitory ones, thereby elevating the two-point discrimination threshold.

\begin{credits}
\subsubsection{\ackname} This research was supported by JSPS KAKENHI Grant-in-Aid for Scientific Research (A) Grant Number 19H01121.
\end{credits}

%
%
%

\bibliographystyle{splncs04}
\bibliography{mybibliography}

\end{document}